# Data-driven investigations of culinary patterns in traditional recipes across the world

Navjot Singh[1,2] and Ganesh Bagler[1*]

[1]*Center for Computational Biology, Indraprastha Institute of Information Technology (IIIT-Delhi), New Delhi, India*
[2]*Delhi Technological University, New Delhi, India*
*Corresponding author: Ganesh Bagler,* `bagler@iiitd.ac.in`

*Abstract*— Cultures around the world have acquired unique culinary practices reflected in traditional recipe compositions. Data-driven analysis has the potential to provide interesting insights into the structure of recipes and organizational principles of cuisines. We provide a curated compilation of 45772 traditional recipes from over 22 regions across the world. Using this resource in conjunction with data of flavor molecules from natural ingredients, we implement data-driven investigations for probing flavor pairing patterns in these recipes. Our analysis reveals non-random recipe compositions characterized with either 'uniform' or 'contrasting' flavor blending and identifies popularity of ingredients as a key contributing factor across all cuisines. Thus we provide a framework for data-driven investigations of culinary patterns in recipes which can be leveraged for applications aimed at food design, generating novel flavor pairings and tweaking recipes for better nutrition and health.

## I. INTRODUCTION

Cooking is a unique endeavor that forms the core of human identity [1]. The knack for combining and processing raw natural ingredients has been suggested to be critical for the evolution of large brains [2]. The increased cognitive capacity has triggered the transformation of *Homo sapiens* into a species endowed with a range of creative faculties including language, arts, technology and abstract science. The penchant for cooking has led to diverse cooking styles across cultures which are crystallized as elaborate procedures in traditional recipes. Passed down from generations to generations, these recipes form the core of cultural identity and shape the dietary habits that are key to diet-linked health problems.

Understandably, cooking has been treated as an artistic endeavor requiring a great deal of culinary intuition acquired through practice. As a consequence, historically the culinary enterprise has not had much scope for organization and integration of knowledge, notwithstanding trends rooted in the molecular basis of food, such as molecular gastronomy [3] and efforts towards understanding the physiology of the taste [4]. While it is clear that preparation and perception of food are complex phenomena that make it difficult to determine underlying rules, search for patterns in recipes can provide much-needed impetus in developing the scientific foundation. Analysis of data pertaining to recipes, ingredients and relevant features can help uncovering underlying patterns in traditional recipes, paving way for innovation in food design and flavor pairings.

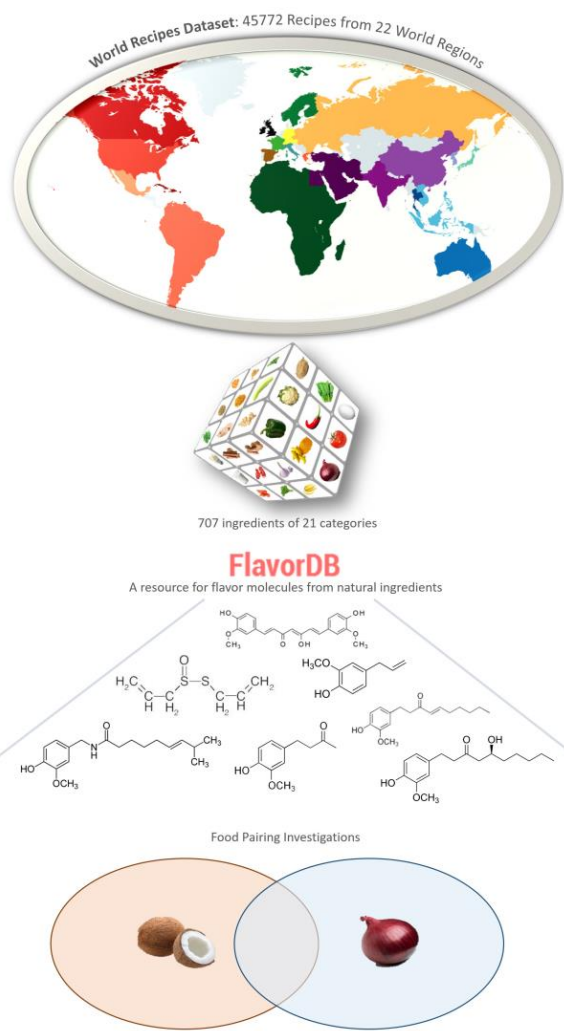

Fig. 1. A strategy for data-driven analysis of flavor pairing patterns in traditional recipes from regions across the world.

Similar to variations in regional languages, cultures across the world have evolved culinary variations. These are encoded in signature ingredient combinations of recipes that characterize a cuisine. In the absence of cultural, climatic and other influences, these recipes would have been composed in a random manner. It has been suggested that traditional culinary practices have a bias towards pairing similar tasting ingredients: ingredients that taste similar go well together [5]. This hypothesis has been, both, confirmed [6] as well as refuted [7], [8] based on empirical evidence from recipes belonging to various world cuisines. It must be highlighted that these contrasting assertions based on the enumeration of food pairing patterns rely on the quality of the underlying data of recipe compositions as well as those of flavor molecules from natural ingredients, apart from the protocol implemented for mapping ingredients in recipes to their natural sources and accounting special ingredients.

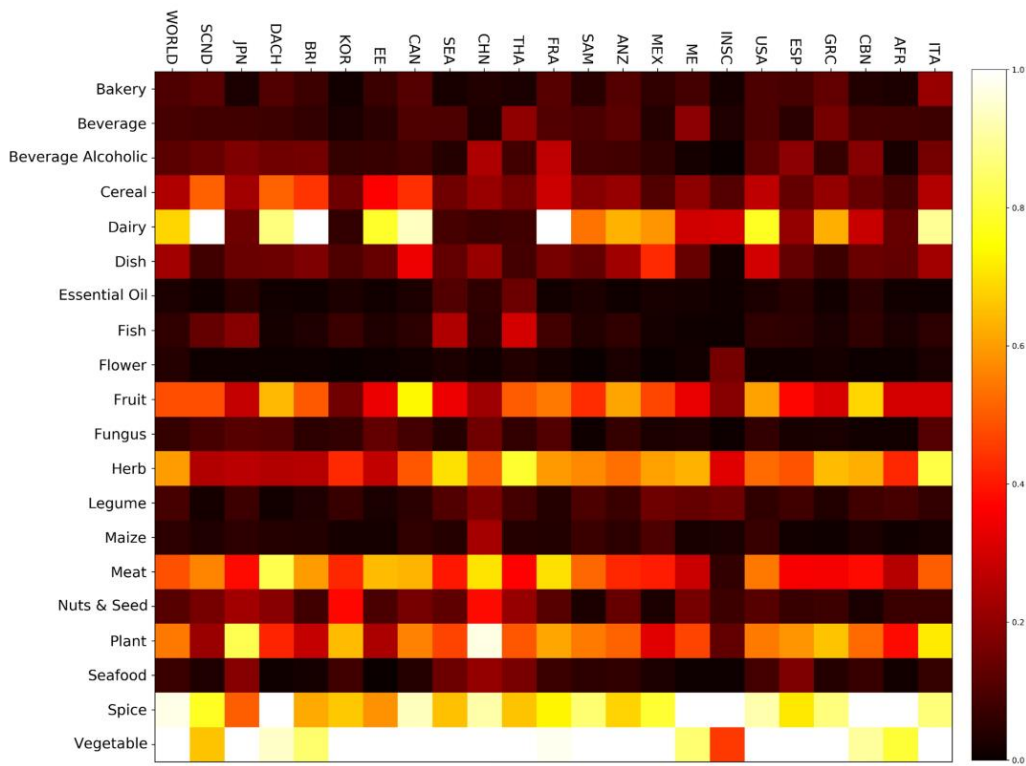

Fig. 2. Compositions of recipes in terms of various ingredient categories.

In light of this, we present a framework for multi-level investigation of traditional recipes rooted in the annotated data of recipes from diverse world cuisines and those of flavor molecules from natural ingredients sourced from FlavorDB [9]. Systematic investigation of deviation patterns from random compositions provide insights into 'culinary fingerprints' [8] that can form the basis for synthesis of novel recipes as well as targeted alterations in existing recipes. In this article, starting with a structured dataset of recipes and flavor molecules in ingredients, we present a strategy for analysis of traditional recipes in search of underlying patterns (Fig. 1).

Our multi-level investigations of recipes from 22 world regions probing patterns in recipe compositions suggest that cuisines across these regions have evolved non-random variations. While some of these regions such as Italy, Africa, Caribbean, Greece, Spain, and USA show strong tendency towards 'uniform food pairing' (blending ingredients of similar flavors), regions such as Scandinavia, Japan, DACH countries, British Isles, Korea, and Eastern European countries showed tendency for 'contrasting food pairing' (blending ingredients of dissimilar flavors). Our analysis presents results contrary to earlier reports [6]–[8], suggesting that quality and completeness of data is a key factor in such studies. Regardless of the inherent pattern, the frequency of use of ingredients emerged as a major factor that influences the food pairing across all cuisines. We also present key ingredients that contribute to the observed culinary patterns. Our study provides curated datasets of recipes from cuisines across world regions and data-driven perspective for their investigation with potential applications for food, flavor, and health.

## II. RESULTS

### A. Multi-level investigation of recipes from across the world

Regional cuisines may be perceived analogous to languages/dialects. The flavor molecules, ingredients, and recipes are for a cuisine, what letters, words, and sentences are for a language. It is important to understand the patterns in the composition of these multi-level building blocks so as to synthesize novel compositions. Data-driven analytics of recipes, ingredients and their constituent flavor molecules provides access to the 'culinary grammar'. With this perspective, we probed diverse world cuisines cutting through all these three levels: Recipe, Ingredient and Flavor Molecules. We compiled data of traditional recipes from diverse regions across the world (see Materials section). Table 1 and Fig. 2 provide statistics of recipes compiled and their unique constituent ingredients belonging to various categories.

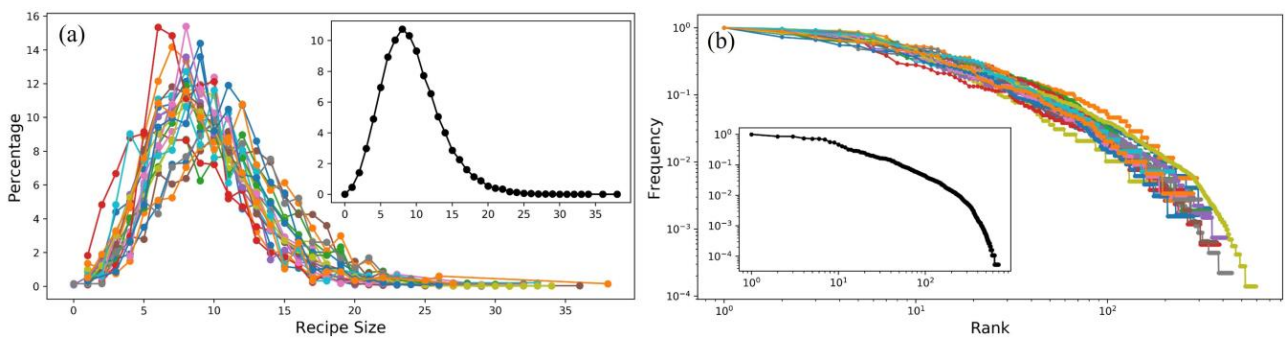

Fig. 3. Statistics of recipe size and popularity of ingredients for all world cuisines from 22 regions. (a) Recipe size distribution and the cumulative statistics (inset). (b) Frequency of use of ingredients, normalized with the most popular ingredient, ranked according to popularity. The inset depicts cumulative statistics. These analyses highlight generic patterns in recipe sizes as well as ingredient popularity across the world cuisines.

| Region (Code) | Recipes | Ingredients |
|---|---|---|
| Africa (AFR) | 651 | 303 |
| Australia & NZ (ANZ) | 494 | 294 |
| British Isles (BRI) | 1075 | 340 |
| Canada (CAN) | 1112 | 368 |
| Caribbean (CBN) | 1103 | 340 |
| China (CHN) | 941 | 302 |
| DACH Countries (DACH) | 487 | 260 |
| Eastern Europe (EE) | 565 | 255 |
| France (FRA) | 2703 | 424 |
| Greece (GRC) | 934 | 280 |
| Indian Subcontinent (INSC) | 4058 | 378 |
| Italy (ITA) | 7504 | 452 |
| Japan (JPN) | 580 | 283 |
| Korea (KOR) | 301 | 198 |
| Mexico (MEX) | 3138 | 376 |
| Middle East (ME) | 993 | 313 |
| Scandinavia (SCND) | 404 | 245 |
| South America (SAM) | 310 | 221 |
| South East Asia (SEA) | 611 | 266 |
| Spain (ESP) | 816 | 312 |
| Thailand (THA) | 667 | 265 |
| USA (USA) | 16118 | 612 |

Table 1. Statistics of recipes and ingredients across world cuisines.

We present a large repertoire of 45772 recipes compiled from various sources and annotated with over 22 world regions (A Database of World Cuisines, Available at http://cosylab.iiitd.edu.in/culinarydb). Each region is well represented, with the lowest number of recipes from Korea (301) and the largest collection of recipes from USA (16118). Among the unique ingredients that were mapped (aliased) to their flavor molecules, the world regions had an average of 321 unique ingredients. These statistics highlight the broad coverage and richness of information in this database. The data of flavor compounds, which are critical for specifying the taste and odor of ingredients by triggering gustatory and olfactory mechanisms, were sourced from FlavorDB, a resource of flavor molecules [9].

We analyzed the recipes for representation across ingredient categories for each cuisine (Fig. 2). We find that at an aggregate level (WORLD), barring the 'Additive' category (data not shown), Vegetable, Spice, Dairy, Herb, Plant, Meat and Fruit categories are used most frequently. However, regional cuisines showed deviations in the preference of popular ingredients. Contrary to the general trend, France, British Isles, and Scandinavia regions use dairy products more prominently than vegetables. Among regions with predominant use of spice were Indian Subcontinent, Africa, Middle East, and Caribbean regions. The heatmap brings out the salient as well as subtle patterns in dietary use of different types of food across world regions.

### B. Patterns in recipes and ingredient popularity

Further, we analyzed the statistics of recipe sizes and ingredient popularity (Fig. 3). Consistent with previous observations [6]–[8], the recipe size distribution is bounded and thin-tailed with an average of nine ingredients per recipe. A typical traditional recipe that has survived over time has an optimal number of ingredients; neither too simple (very few ingredients) nor overloaded with ingredients, which would make it difficult to cook and propagate. When probed for pattern in ingredient popularity, again consistent with previous observations [6]–[8] all the world cuisines showed an exceptionally consistent scaling phenomenon. Every region has its special ingredients that are most popular and dominate the cuisine with excessive use in cooking preparations. Nonetheless, the pattern of ingredients ranking shows an exceptionally consistent pattern across cuisines.

### C. Patterns in flavor pairing

Going beyond the layers of recipe and ingredients, we further incorporated the well-curated data of flavor molecules in natural ingredients from FlavorDB [9] to establish a framework for the analysis of flavor pairing patterns across world cuisines (Fig. 1). The flavor pairing (also known as food pairing) enumerates the extent of overlap in flavor profiles of constituent ingredients in recipes. When compared with the randomized cuisine that follows the recipe size distribution of its traditional counterpart and incorporates exactly the same set of ingredients, a cuisine may be characterized with (a) uniform blend of ingredients, (b) contrasting blend of ingredients, or (c) could be indistinguishable from the random cuisine. Fig. 4 depicts the Z-score (See Methodology section) for each of the 22 world cuisines investigated, suggesting the statistical significance of deviation from its random counterpart.

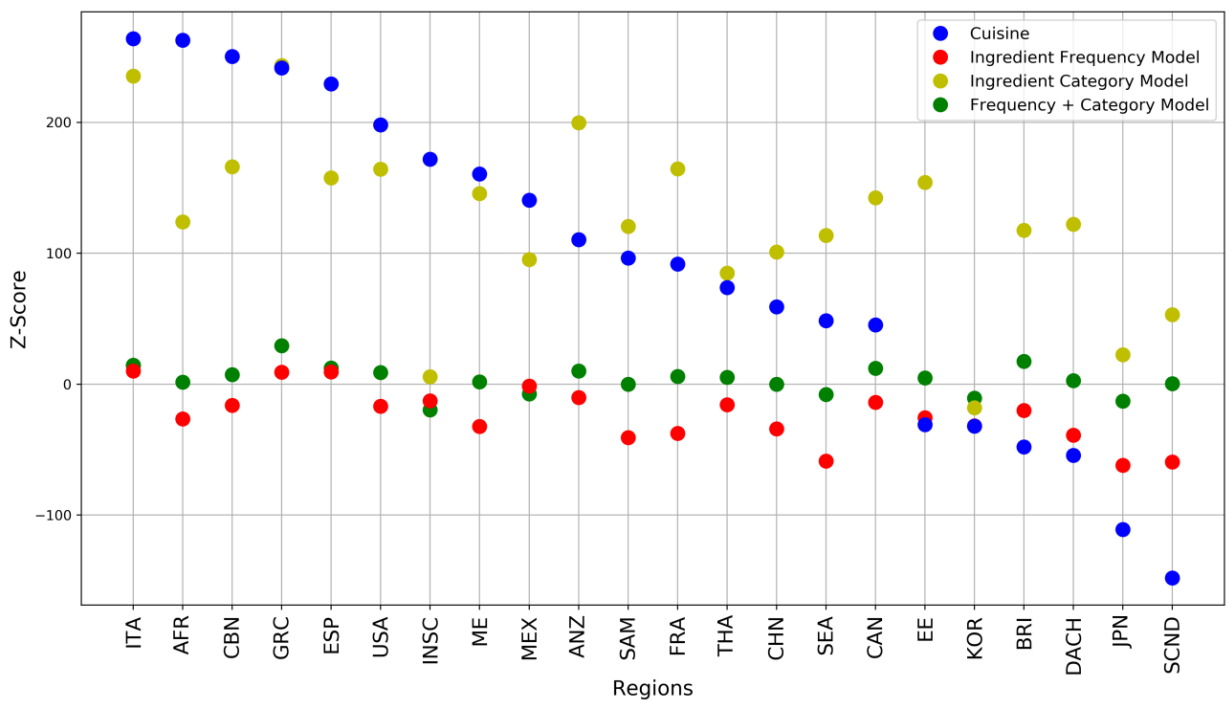

Fig. 4. Food pairing analysis of cuisines from 22 world regions based on data of world cuisines and flavor compounds from natural ingredients. Compared to their randomized counterparts, every cuisine showed deviation in its ingredient pairing pattern towards either uniform blend of ingredients (positive food pairing) or contrasting blend of ingredients (negative food pairing). Preserving the observed frequency of use of ingredients reproduces the food pairing patterns to a large extent. The model emulating preservation of category composition was unable to reproduce the original food pairing.

Interestingly, cuisines from 16 world regions showed uniform food pairing (positive food pairing) suggesting a tendency of their recipes to mix ingredients of similar flavor profiles, consistent with the hypothesis suggested by Chef Heston Blumenthal [5], [6]. These cuisines, belonging to diverse geographic and climatic regions, were spread across continents: Italy, Africa, Caribbean, Greece, Spain, USA, Indian Subcontinent, Middle East, Mexico, Australia & New Zealand, South America, France, Thailand, China, South East Asia and Canada. The remaining 6 regions showed a tendency towards contrasting food pairing (negative food pairing): Scandinavia, Japan, DACH Countries, British Isles, Korea, and Eastern Europe. It should be noted that some of these results are contrary to observations made earlier [6]–[8]. At the same time, interestingly none of the cuisines shows food pairing that is indistinguishable from its random counterpart.

We surmise that traditional recipes from diverse cultures have evolved ingredient compositions deviating from random mixing to acquire either uniform blends or contrasting blends. It would be interesting to study whether such a trend reflects a basis for palatability of recipes and its implications for nutrition/health.

We implemented three models by incorporating the ingredient popularity, ingredient category composition of recipes and both, to assess their contribution to the observed food pairing patterns. Interestingly, we find that ingredient popularity accounts for both the positive as well as negative food pairing patterns across all cuisines. The ingredient category composition of the recipes, on the other hand, are not critical for food pairing. We also identified ingredients contributing to the observed food pairing pattern in all the cuisines. Fig. 5 depicts the top 3 ingredients that contribute the most to cuisines observed to follow positive (Fig. 5(a)) and negative (Fig. 5(b)) food pairing.

III. MATERIALS

Our analysis involved three levels of information pertaining to traditional recipes, namely, recipes, ingredients and flavor molecules. All data are available at http://cosylab.iiitd.edu.in/culinarydb (A Database of World Cuisines) and FlavorDB (http://cosylab.iiitd.edu.in/flavordb).

*A. Recipes*

A total of 45772 recipes were obtained from various sources: AllRecipes (https://www.allrecipes.com) (16177), Food Network (https://www.foodnetwork.com) (15917), Epicurious (https://www.epicurious.com) (11069) and TarlaDalal (https://www.tarladalal.com) (2609). For each recipe, details such as recipe name, list of ingredients and cooking procedure were extracted. For the purpose of food pairing analysis, each recipe was treated as an unordered list of ingredients. Therefore, only those recipes were considered for which information of cuisine and ingredients list were available.

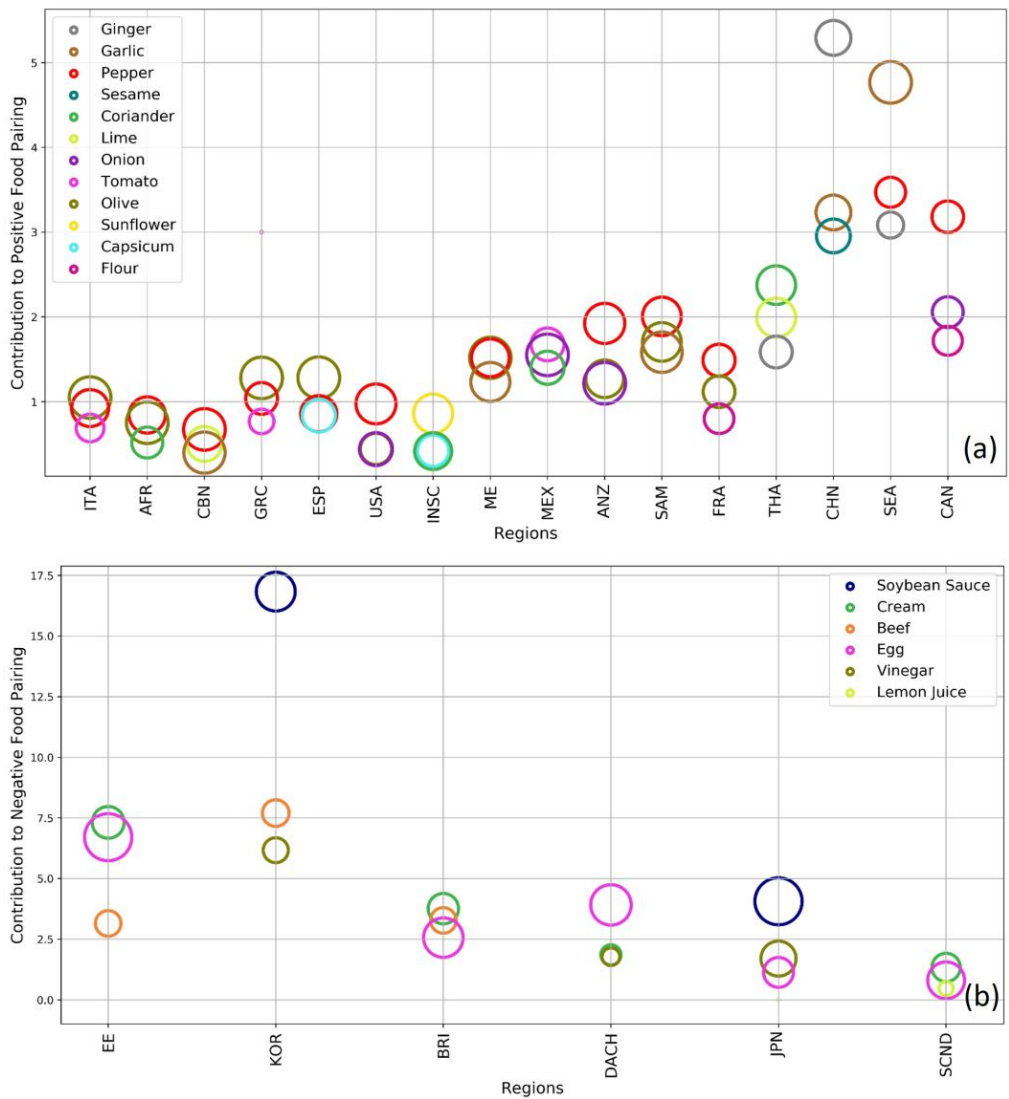

Fig. 5. Top 3 ingredients contributing to the (a) positive and (b) negative food pairing to cuisines with the dominance of uniform or contrasting blend, respectively.

We integrated recipes from all the sources and grouped them in 22 distinct geo-cultural 'regions' while ensuring that each region has enough recipes attributed to it. Please refer to Table 1 for the lists regions. Due to an inadequate number of recipes, while 207 recipes from Portugal, Belgium, Central America, and Netherlands were used for aggregate analysis (WORLD), they were not considered as independent regions.

### B. Ingredients

Ingredient list from FlavorDB was used as a primary source of ingredient information. However, this list was tailored for the analysis of recipes. 29 generic and noisy entities were removed. For certain ingredients, their common names were added. For example, corresponding to bread, beer, and yogurt, their synonyms bun, lager, and curd were included. Also, linguistic and spelling variations of ingredients were accounted for. For example, whiskey and whisky, asafoetida and *hing*, chili and chile etc. were considered as the same entity.

Further, 13 new ingredients were added to the list curated from FlavorDB. These were specific ingredients for which flavor profiles were available and were coarse-grained in FlavorDB, but are important due to their role in recipes: anise oil, apple juice, coconut milk, coconut oil, hops bear, lemon juice, brown rice, tomato juice, tomato paste, tomato puree, coriander seed, pork fat, and cured-ham. Data from Ahn *et al.* [6], was used to include 4 more unique ingredients: cayenne, yeast, tequila, and sauerkraut. Information for 7 additives was manually added considering their high usage: baking powder, monosodium glutamate, citric acid, cooking spray, gelatin, food coloring and liquid smoke. For the last four additives, no flavor profile was added. In addition to 840 unique basic ingredients, we compiled a list of 103 'compound ingredients'. This list includes readymade spice combinations, sauces and other commonly used dishes which are often considered as core ingredients themselves. For example 'half half' consists of milk and cream, 'mayonnaise' is made up of oil, egg and lemon juice. Also, to account for limited number of flavor molecules in certain ingredients, some entities were bundled to create a compound ingredient (such as black bear, polar bear and brown bear were combined to form 'bear').

Each ingredient was classified into one of the 21 categories: Vegetable, Dairy, Legume, Maize, Cereal, Meat, Nuts and Seeds, Plant, Fish, Seafood, Spice, Bakery, Beverage Alcoholic, Beverage, Essential Oil, Flower, Fruit, Fungus, Herb, Additive, and Dish.

*C. Flavor Molecules*

FlavorDB [9] (http://cosylab.iiitd.edu.in/flavordb) was used as the source of flavor molecules from natural ingredients, in addition to obtaining details from other sources as mentioned above. The flavor profile of an ingredient refers to repertoire of empirically reported flavor molecules. For 'compound ingredients', the flavor profile was generated by creating a list of unique flavor molecules after pooling flavor molecules of its constituent ingredients.

IV. METHODOLOGY

*A. Ingredient Aliasing*

After obtaining a curated list of recipes, ingredients and their flavor profiles, the next task was to create a mapping between ingredients used in recipes and ingredient list complete with their flavor profile. The process of aliasing enables flavor pairing analysis. The phrases containing ingredient names (For eg: '2 jalapeno peppers, roasted and slit') were processed through a multi-step protocol. To begin with, they were converted to lower case, and the punctuation marks, special characters, and stopwords, including some culinary stopwords, were removed. Further, all words in the phrases were converted into their singular forms. These steps were implemented with the help of NLTK (Natural Language Toolkit) and 'inflect' python packages.

This protocol involved robust string processing to take into account variations in writing ingredient spellings. While maximizing the information retrieval from cooking procedures, care was taken to minimize the false positives. Partial matches and unrecognized ingredients were explicitly labeled for manual curation. N-grams (up to 6-grams) were created on the basis of partial and unrecognized ingredients to identify commonly occurring ingredients which were either not present in the database or were variations of existing entities.

*B. Food Pairing*

Food pairing is measured in terms of overlap of flavor molecules for a pair of ingredients. It enumerates the recipe composition pattern by accounting similarity in flavor profiles of its constituent ingredients. The food pairing hypothesis suggests that ingredients sharing flavor compounds are more likely to taste well together than ingredients that do not. We quantified the food pairing pattern in 22 cuisines belonging to different world regions.

The concept of food pairing is rooted in the premise that flavor molecules are one of the primary causal elements involved in food sensation. For a given recipe $R$, the interrelationship among its ingredients by virtue of shared flavor compounds can be quantified with the help of the food pairing score which is defined as follows: $N_s^R$ for a recipe $R$ with $n$ ingredients can be defined as:

$$N_s^R = \frac{2}{n(n-1)} \sum_{i,j \in R, i \neq j} |F_i \cap F_j|$$

Where $F_i$ represents the flavor profile (the set of flavor compounds) of an ingredient $i$ and $n$ is the number of ingredients in the recipe (recipe size).

Average flavor sharing of a cuisine $C$ with $N_s^C$ recipes can be calculated by averaging the food pairing score of all its constituent $n_r$ recipes:

$$N_s^C = \frac{1}{n_r} \sum_R N_s^R$$

For each cuisine, this average flavor sharing score was compared with its corresponding randomized cuisine to assess its statistical relevance by computing:

$$\Delta N_s^C = N_s^C - N_s^{rand}$$

Four types of models of randomized cuisines were created to identify factors that account for food pairing patterns. In each of these, the exact set of ingredients and the recipe size distribution from the original cuisine were preserved.

*Random Cuisine*: Ingredients were chosen uniformly from the set of ingredients used in a particular cuisine for which randomized cuisine was being created.

*Ingredient Frequency Model*: A frequency-preserved model was generated in which the frequency of use of ingredients in the respective cuisine was preserved.

*Ingredient Category Model*: A category-preserved control was generated in which the category composition of the recipe was preserved; ingredients were randomly chosen from each constituent category.

*Frequency + Category Model*: A composite frequency-and-category preserved randomised control was created where the category composition was maintained and each ingredient was chosen with probability consistent with its frequency in the respective original cuisine.

100,000 recipes were generated for the random control and models. The statistical significance of $\Delta N_s^C$ was measured with the help of a Z-score:

$$Z = \sqrt{n_r^{rand}} \frac{N_s^C - N_s^{rand}}{\sigma_{rand}}$$

Where $n_r^{rand}$ denotes the number of recipes in the random cuisine (100,000) and $\sigma_{rand}$ denotes the standard deviation. The food pairing of a Cuisine is represented by the $Z$-score computed in comparison with the Random Cuisine. It enumerates the extent to which the food pairing in randomized cuisine differs from that of the actual cuisine.

*C. Ingredient Contribution*

The ingredient contribution $\chi_i$, of each ingredient $i$ to the flavor sharing of a cuisine, quantifies the extent to which its presence affects the magnitude of $N_s^C$. It was measured as 'percentage change in food pairing score' in response to removal of the ingredient from the cuisine.

## V. CONCLUSIONS

Data-driven analysis can provide interesting insights into the structure of recipes and organizational principles of cuisines. Such a quantitative approach for the study of patterns that dominate world cuisines enables search for underlying rules that shape our culinary habits. Our analysis suggests that traditional recipes from various cuisines tend to deviate from the food pairing pattern expected in randomized recipes. Apparently, as the cuisines evolve to acquire a distinct character, their recipes develop non-random ingredient combination leaning towards either uniform or contrasting blend of flavors. While it is plausible that various constraints (geographic, climatic, cultural) may play a role in the evolution of such patterns, a simple copy-mutate model has been shown to explain such patterns [10]. Our studies also reveal that the ingredient popularity to a large extent accounts for the food pairing patterns. On the other hand, preserving the category composition itself is not sufficient to reproduce the original food pairing.

Apart from paving the way for data driven investigations of recipes, our studies open up many questions. How robust are the patterns to changes in recipes data and flavor profiles? What are the patterns at higher order n-tuples (i.e. instead of pairs what if one were to compute triples and quadruples of ingredients)? How to incorporate details of recipe preparation and quantity of ingredients to take the analysis closer to the reality? How to incorporate transformation of flavor in the process of cooking? What strategies could be developed to generate novel recipes that are palatable and healthier? Could it be possible to enumerate the taste of a recipe? How to use culinary fingerprints for designing novel food and beverages?

We believe that studies aimed at answering such questions are establishing the foundations of 'Computational Gastronomy' and would help transform the landscape of food making it possible to leverage it for better health and nutrition.

## ACKNOWLEDGMENT

G.B. thanks the Indraprastha Institute of Information Technology (IIIT-Delhi) for providing computational facilities and support. N.S. is a Research Intern in Dr. Bagler's lab (Complex Systems Laboratory) at the Center for Computational Biology, and is thankful to IIIT-Delhi for the support.